\documentclass[preprint,showpacs,preprintnumbers,amsmath,amssymb]{revtex4}
\usepackage{booktabs}
\usepackage{mathrsfs}
\usepackage{epsfig}
\usepackage{graphicx}
\usepackage{dcolumn}
\usepackage{bm}
\usepackage{amsmath}
\usepackage{slashed}       
\usepackage{amssymb}
\usepackage{amsfonts}
\usepackage{multirow}

\let\jnfont=\rm
\def\NPB#1,{{\jnfont Nucl.\ Phys.\ B }{\bf #1},}
\def\PLB#1,{{\jnfont Phys.\ Lett.\ B }{\bf #1},}
\def\EPJC#1,{{\jnfont Eur.\ Phys.\ Jour.\ C }{\bf #1},}
\def\PRD#1,{{\jnfont Phys.\ Rev.\ D }{\bf #1},}
\def\PRL#1,{{\jnfont Phys.\ Rev.\ Lett.\ }{\bf #1},}
\def\MPLA#1,{{\jnfont Mod.\ Phys.\ Lett.\ A }{\bf #1},}
\def\JPG#1,{{\jnfont J.\ Phys.\ G}{\bf #1},}
\def\CTP#1,{{\jnfont Commun.\ Theor.\ Phys.\ }{\bf #1},}
\def\ZPC#1,{{\jnfont Z.\ Phys.\ C }{\bf #1},}
\def\JHEP#1,{{\jnfont JHEP \ }{\bf #1},}
\def\lsim{\raise0.3ex\hbox{$<$\kern-0.75em\raise-1.1ex\hbox{$\sim$}}}
\def\gsim{\raise0.3ex\hbox{$>$\kern-0.75em\raise-1.1ex\hbox{$\sim$}}}

\begin{document}
\preprint{\parbox{1.2in}{\noindent arXiv:1309.4939}}

\title{A light Higgs scalar in the NMSSM confronted with \\ the latest LHC Higgs data}

\author{Junjie Cao$^{1,2}$,  Fangfang Ding$^1$, Chengcheng Han$^3$, Jin Min Yang$^3$, Jingya Zhu$^3$}

\affiliation{
  $^1$  Department of Physics,
        Henan Normal University, Xinxiang 453007, China \\
  $^2$ Center for High Energy Physics, Peking University,
       Beijing 100871, China \\
  $^3$ State Key Laboratory of Theoretical Physics,
      Institute of Theoretical Physics, Academia Sinica, Beijing 100190, China
      \vspace{1cm}}

\begin{abstract}
In the Next-to-Minimal Supersymemtric Standard Model (NMSSM), one of the neutral Higgs scalars
(CP-even or CP-odd) may be lighter than half of the SM-like Higgs boson.
In this case, the SM-like Higgs boson $h$ can decay into such a light scalar pair and consequently
the $\gamma \gamma$ and $Z Z^\ast$ signal rates at the LHC will be suppressed.
In this work, we examine the constraints of the latest LHC Higgs data on such a possibility.
We perform a comprehensive scan over the parameter space of the NMSSM by considering various
experimental constraints and find that the LHC Higgs data can readily constrain the
parameter space and the properties of the light scalar, e.g., at $3\sigma$ level this
light scalar should be highly singlet dominant and
the branching ratio of the SM-like Higgs boson decay into the scalar pair should be less than about $30\%$.
Also we investigate the detection of this scalar at various colliders.
Through a detailed Monte Carlo simulation we find that
under the constraints of the current Higgs data this light scalar can be accessible at
the LHC-14 with an integrated luminosity over 300 fb$^{-1}$.
\end{abstract}


\maketitle

\section{Introduction}

The existence of a new scalar has been discovered by the ATLAS and CMS collaborations
with a significance of $9 \sigma$ and $7\sigma$, respectively
\cite{1207-a-com,1207-c-com,1303-atlas-Moriond,1303-c-com}.
So far the mass of this scalar is rather precisely determined to be around 125 GeV,
and its other properties, albeit with large experimental uncertainties, agree with
the Standard Model (SM) prediction \cite{1303-c-com,1303-a-com}.
In spite of this, this newly discovered scalar has been interpreted in various
new physic models since the SM has the gauge hierarchy problem and cannot provide
a dark matter candidate.
The studies in this direction have been carried out intensively in
low energy supersymmetric models and the NMSSM was found to be most
favorved \cite{125-cao, 125-mssm, 125-NMSSM, susy-our, susy-status, cao-cmssm, 125-cmssm, nMSSM}.

In this work we focus on the NMSSM, which is the simplest extension of the MSSM
with one extra gauge singlet Higgs field \cite{NMSSM}.
One virtue of such an extension is that it provides a dynamical mechanism for the
generation of the parameter $\mu$ and thus solves the so-called
$\mu$-problem suffered by the MSSM \cite{MuProblem}.
Another virtue is that the interactions of the singlet field in the Higgs sector give a new contribution
to the tree-level mass of the SM-like Higgs boson and thus alleviate the little hierarchy problem
\cite{NMSSM-FT, susy-status}.
For the LHC phenomenology, one notable feature of the NMSSM is that a Higgs scalar (CP-even or CP-odd)
may be rather light \cite{lighta-very, lighta-hc},
which can affect the signals of the sparticles at the LHC \cite{lighta-MC, NMSSM-Neu}.
For example, if the lightest supersymmetric particle is singlino-like, squarks may decay
dominantly as  \cite{NMSSM-Neu}
$\tilde{q} \to q \tilde{\chi}_{2,3}^0 \to q \tilde{\chi}_1^0 S \to q \tilde{\chi}_1^0 b \bar{b}$,
where $\tilde{\chi}_{2}^0$ and $\tilde{\chi}_{3}^0$ represent the second and the third lightest
neutralino respectively, and $S$ denotes a light scalar.

We note that, if this scalar is lighter than half of the SM-like Higgs boson,
the SM-like Higgs boson can decay exotically into the light scalar pair
\cite{NMSSM-lightDM, lighta-hsm, lighta-hsm-125}.
Since the width of the Higgs boson in the SM is quite narrow (about 4 MeV),
such an exotic decay may have a sizable branching ratio.
This in return can suppress greatly the visible signals
of the SM-like Higgs boson at the LHC.
Motivated by this observation, we in this work investigate the constraints
of the latest LHC Higgs data on the properties of such a light scalar.
We will also study  the detection of this scalar at the LHC-14 via a
detailed Monte Carlo simulation.

The outline of this paper is as follows. In Section II we briefly review the NMSSM model.
Then in Section III we scan the parameter space of the NMSSM under current experimental constraints.
In Section IV the properties of the light scalar are analyzed and its detection at the LHC-14 is
studied via a detailed Monte Carlo simulation.
Finally, we present our conclusion in Section V.

\section{The Higgs sector of the NMSSM }
As one of the most economical extensions of the MSSM, the NMSSM contains two SU(2) doublet Higgs fields
and one gauge singlet Higgs field \cite{NMSSM}. Traditionally, these fields are labeled by
\begin{eqnarray}
\hat{H}_{u}=\left(        \begin{array}{c}
          H^{+}_{\rm u} \\
          v_{u}+ \frac{\phi_{u}+i\varphi_{u}}{\sqrt{2}} \\
        \end{array} \right) ,
\hat{H}_{d}=\left(         \begin{array}{c}
          v_{d}+ \frac{\phi_{d}+i\varphi_{d}}{\sqrt{2}} \\
          H^{-}_{\rm d} \\
         \end{array}        \right)   ,
\hat{S}=v_{s}+ \frac{\phi_{s}+i\varphi_{s}}{\sqrt{2}},  \label{H-F}
\end{eqnarray}
where $H^+_i$, $\phi_i$ and $\varphi_i$ ($i=u,d$) represent the charged, neutral CP-even and neutral CP-odd
component fields respectively, and $v_u$, $v_d$ and $v_s$  are the vacuum expectation values with
$v_u/v_d=\tan\beta$ and $\sqrt{v_u^2+v_d^2}=v\equiv174~{\rm GeV}$. Since one purpose of the extension
is to solve the $\mu$-problem of the MSSM, a $Z_3$ symmetry is implemented in the construction of the
superpotential to avoid the appearance of parameters with mass dimension.  Consequently,  the
superpotential and the soft breaking  terms  in the NMSSM are given by \cite{NMSSM}
\begin{eqnarray}
  W^{\rm NMSSM} &=& W_F + \lambda\hat{H_u} \cdot \hat{H_d} \hat{S}
  +\frac{1}{3}\kappa \hat{S^3},\\
  V^{\rm NMSSM}_{\rm soft} &=& \tilde m_u^2|H_u|^2 + \tilde m_d^2|H_d|^2
  +\tilde m_S^2|S|^2 +( \lambda A_{\lambda} SH_u\cdot H_d
  +\frac{1}{3}\kappa A_{\kappa} S^3 + h.c.),
\end{eqnarray}
where $\hat{H_u}$, $\hat{H_d}$ and $\hat{S}$ are Higgs superfields, $W_F$ is the superpotential
of the MSSM without the $\mu$-term, and $\tilde{m}_{u}$, $\tilde{m}_{d}$, $\tilde{m}_{S}$, $A_\lambda$
and $A_\kappa$ are soft-breaking parameters.

In order to present the mass matrices of the Higgs fields in a physical way,
we redefine the Higgs fields as \cite{hmass-NMSSM}
\begin{eqnarray}
H_1=\cos\beta H_u -\varepsilon \sin\beta H_d^*, ~~
H_2=\sin\beta H_u +\varepsilon \cos\beta H_d^*, ~~H_3 = S,
\end{eqnarray}
where $\varepsilon_{12}=\varepsilon_{21}=-1$ and $\varepsilon_{11}=\varepsilon_{22}=0$.
With such a definition, $H_i$ ($i=1,2,3$) are given by
\begin{eqnarray}
H_1 = \left ( \begin{array}{c} H^+ \\
       \frac{S_1 + i P_1}{\sqrt{2}}
        \end{array} \right),~~
H_2 & =& \left ( \begin{array}{c} G^+
            \\ v + \frac{ S_2 + i G^0}{\sqrt{2}}
            \end{array} \right),~~
H_3  = v_s +\frac{1}{\sqrt{2}} \left(  S_3 + i P_2 \right),
\label{fields}
\end{eqnarray}
where $\phi_{s}$ and $\varphi_{s}$ in Eq.(\ref{H-F}) are rewritten as $S_3$ and $P_2$ respectively.
Obviously, the field $H_2$ corresponds to the SM Higgs field with $G^+$ and $G^0$ denoting Goldstone
bosons, and $S_2$ representing the SM Higgs boson.

In the CP-conserving NMSSM, the fields $S_1$, $S_2$ and $S_3$ mix to form three (instead of two in the MSSM)
physical CP-even Higgs bosons $h_i$ ($i=1,2,3$). In the basis ($S_1$, $S_2$, $S_3$), the elements of the
corresponding mass matrix are given by \cite{hmass-NMSSM}
\begin{eqnarray}
  M^2_{11} &=&  M^2_A + (m^2_Z -\lambda^2 v^2) \sin^2 2\beta, \nonumber \\
  M^2_{12} &=&  -\frac{1}{2}(m^2_Z-\lambda^2 v^2)\sin4\beta, \nonumber \\
  M^2_{13} &=&  -(\frac{M^2_A}{2\mu/\sin2\beta}+\kappa v_s) \lambda v\cos2\beta, \nonumber \\
  M^2_{22} &=&  m_Z^2\cos^2 2\beta +\lambda^2v^2\sin^2 2\beta, \nonumber \\
  M^2_{23} &=&  2\lambda\mu v[1-(\frac{M_A}{2\mu/\sin2\beta})^2 -\frac{\kappa}{2\lambda}\sin2\beta], \nonumber \\
  M^2_{33} &=&  \frac{1}{4}\lambda^2 v^2(\frac{M_A}{\mu/\sin2\beta})^2 +\kappa v_s A_{\kappa}+4(\kappa v_s)^2 -\frac{1}{2}\lambda\kappa v^2 \sin 2\beta.
  \label{MS}
\end{eqnarray}

Similarly,  the fields $P_1$ and $P_2$ mix to form two physical CP-odd Higgs bosons $A_i$ ($i=1,2$), and
in the basis ($P_1$, $P_2$) the mass matrix elements for CP-odd Higgs sector are given by
\begin{eqnarray}
  M^2_{-11} &=& M^2_A
  =\frac{2\mu}{\sin2\beta}(A_{\lambda}+\kappa v_s), \nonumber \\
  M^2_{-22} &=& M^2_P
  =\lambda^2 v^2 (\frac{M_A}{2\mu/\sin2\beta})^2
  +\frac{3}{2}\lambda\kappa v^2\sin2\beta -3 \kappa v_s A_{\kappa},  \nonumber \\
  M^2_{-12} &=&
  \lambda v \frac{M_A^2}{2\mu/\sin2\beta} -3\lambda \kappa v_s v. \label{MA}
\end{eqnarray}

About the Higgs sector of the NMSSM, the following points should be noted:
\begin{itemize}
\item Compared with the MSSM where only two parameters are involved in the Higgs sector,
six parameters are needed to describe the Higgs sector of the NMSSM \cite{NMSSM}.
These parameters are usually chosen as
\begin{eqnarray}
\lambda, \quad \kappa,
\quad \tan \beta=\frac{v_u}{v_d},
\quad \mu = \lambda v_s,
\quad M_A^2= \frac{2 \mu}{\sin 2 \beta}(A_\lambda + \kappa v_s),
\quad M_P.
\end{eqnarray}
Since the NMSSM predicts one more CP-odd Higgs field than the MSSM, $M_A$ here no longer represents the
mass of one CP-odd state. Obviously, the Higgs sector of the NMSSM is quite complicated.
\item After diagonalizing the mass matrix in Eq.(\ref{MS}), one can get the mass eigenstates of CP-even states
$h_i$  as
\begin{eqnarray}
h_i = \sum_{j=1}^3 V_{ij} S_j, \nonumber
\end{eqnarray}
where $V_{ij}$ is the element of the transition matrix satisfying $V_{i1}^2 + V_{i2}^2 + V_{i3}^2 = 1$,
and it represents the component of $S_j$ in the physical state $h_i$.
In the following, we assume $m_{h_3} > m_{h_2} > m_{h_1}$,
and call the state whose squared component coefficient
of $S_2$ larger than 0.5 the SM-like Higgs boson.

Similarly, the mass eigenstates of the CP-odd states $A_i$  are given by
\begin{eqnarray}
A_i = \sum_{j=1}^2 U_{ij} P_j. \nonumber
\end{eqnarray}
If the lighter state $A_1$ satisfies $U_{11}^2 > 0.5$, we call it doublet dominated;
otherwise we call it singlet dominated.
\item Like the MSSM, the mass of the SM-like Higgs boson may be greatly changed by the radiative corrections.
Denoting the loop-corrected mass matrix of the CP-even states by $\tilde{M}^2$,  one can conclude that
for $\tilde{M}_{11}^2 > \tilde{M}_{33}^2 > \tilde{M}_{22}^2$, the state $h_1$ corresponds to the SM-like Higgs boson,
while for $\tilde{M}_{11}^2 > \tilde{M}_{22}^2 > \tilde{M}_{33}^2$, the state $h_2$ is the SM-like Higgs boson \cite{125-cao}.
\item Obviously, in order to get a light CP-odd Higgs boson, either $M_A$ or $M_P$ should be moderately small,
and a large $M_{-12}^2$ can further suppress the mass of the lighter CP-odd state.
\end{itemize}

\section{Numerical result and discussion}
In this work, we first perform a comprehensive scan over the parameter space of
the NMSSM by considering various experimental constraints. Then for the surviving samples
we investigate the features of the light scalar.
Since for the NMSSM there are too many free
parameters, we make the following assumptions to simplify our analysis:
\begin{itemize}
\item First, we note that the first two generation squarks have little effects on the Higgs sector of
the NMSSM, and the LHC search for SUSY particles implies that they should be
heavier than 1 TeV. So we fix all soft breaking parameters (i.e. soft masses and
trilinear coefficients) in this sector to be 2 TeV. We checked that our conclusions are
not sensitive to this sector.
\item Second, considering that the third generation squarks can change significantly
the properties of the Higgs bosons, we set free all soft parameters in this sector
except that we assume $m_{U_3}=m_{D_3}$ and $A_t = A_b$ to reduce the number of free parameters.
\item Third, since we require the NMSSM to explain the discrepancy of
the measured value of the muon anomalous magnetic moment from its SM prediction,
i.e., $a_\mu^{exp} - a_\mu^{SM}= (28.7 \pm 8.0 ) \times 10^{-10}$ \cite{PDG2012}, we assume all soft
breaking parameters in the slepton sector to take a common value $m_{\tilde{l}}$ and treat $m_{\tilde{l}}$
as a free parameter.
\item Finally, we note that our results are  not sensitive to gluino mass, we fix it at 2 TeV.
We also assume the grand unification relation $3 M_1/5 \alpha_1 = M_2/\alpha_2$ for
electroweak gaugino masses.
\end{itemize}

With above assumptions, we use the package NMSSMTools-4.0.0 \cite{NMSSMTools} to scan randomly the free parameters
of the model in the following ranges
\begin{eqnarray}\label{NMSSM-scan}
&& 0.1 \leq \lambda, \kappa \leq 0.8,
~~~1{\rm ~GeV}\leq M_A, M_P\leq 2 {\rm ~TeV},
\nonumber\\
&& 1\leq\tan\beta \leq 30,
~~~100{\rm~GeV}\leq \mu, M_2, m_{\tilde{l}} \leq 1 {\rm ~TeV},
\nonumber\\
&& |A_{t}|\leq 5 {\rm ~TeV},
~~~100{\rm ~GeV}\leq M_{Q_3},M_{U_3} \leq 2 {\rm ~TeV}.
\end{eqnarray}
In our scan, we only keep the samples that predict a SM-like Higgs boson $h$ with mass around 125 GeV
(e.g. $123 {\rm GeV} \leq m_h \leq 127 {\rm GeV}$) along with a light neutral Higgs scalar
(CP-even or CP-odd) with mass less than $m_h/2$, and meanwhile satisfy the following constraints:
\begin{itemize}
\item[(1)] All the constraints implemented in the package NMSSMTools-4.0.0.
These constraints are from the vacuum stability, the LEP search for sparticles
(including lower bounds on various sparticle masses, the upper bounds on the
neutralino pair production rates), the $Z$-boson invisible decay,
the $\Upsilon$ decay into a light scalar plus one photon \cite{BaBar-LowH}, the $B$-physics observables
(such as the branching ratios for $B \to X_s \gamma$, 
$B_s \to \mu^+ \mu^-$ and $B^+ \to \tau^+ \nu_\tau$, 
and the mass differences $\Delta M_d$ and $\Delta M_s$) 
\cite{BaBar-Bph, LHCb-BsMuMu, PDG2012},
the discrepancy of the muon anomalous magnetic moment,
the dark matter relic density \cite{Planck2013} and the XENON100(2012)
limits on the scattering rate of dark matter with nucleon \cite{XENON2012, cao-dm}.
In imposing the constraint from a certain observable which
has an experimental central value, we use its latest measured result and require
the NMSSM to explain the result at $2\sigma$ level.

\item[(2)] The constraints from the search for Higgs bosons at the LEP, the Tevatron and the LHC.
We implement these constraints by the package HiggsBounds-4.0.0 \cite{HiggsBounds}.

\item[(3)] Indirect constraints from electroweak precision
observables such as $\rho_{\ell}$, $\sin^2 \theta_{eff}^{\ell}$ and
$M_W$, or their combinations $\epsilon_i (i=1,2,3)$ \cite{Altarelli}.
We require $\epsilon_i$ to be compatible with the
LEP/SLD data at $95\%$ confidence level \cite{LEP-Report}. We also
require $R_b$ in the NMSSM is within the $2 \sigma$ range of
its experimental value.
We compute these observables with the formula presented in \cite{cao-zbb}.
\end{itemize}

For each surviving sample, we further perform a fit using the latest Higgs data presented at the
Rencontres de Moriond 2013.
These data include the measured signal strengthes for
$\gamma \gamma$, $ZZ^\ast$, $W W^\ast$, $b\bar{b}$ and $\tau \bar{\tau}$ channels,
and their explicit values are summarized in Fig.2 of \cite{1303-a-com} for the ATLAS results,
in Fig.4 of \cite{1303-c-com} for the CMS results and in Fig.15 of \cite{1303-t-com}
for the CDF+D0 results.
We totally use 24 sets of experimental data with 22 of them corresponding
to the measured signal strengthes and the other 2 being the combined mass of the Higgs
boson reported by the ATLAS and the CMS collaborations respectively.
As in our previous works \cite{hfit-our},
we use the method first introduced in \cite{hfit-first} to perform the fit,
and properly consider the correlations of the data as in \cite{1212-Gunion, 1307-Gunion}.
As will be shown below, the $\chi^2$ values in the fit vary from several tens to 170 for
the surviving samples of the scan, and in optimal case it may be as low as about $17$. In
our discussion, we will pay particular attention to  the surviving samples with $\chi^2 \leq 26$. These samples
can be used to get the $3 \sigma$ range of any observable $O_i$ once they are projected on
the $O_i$ versus $\delta \chi^2$ plane, so hereafter we call them $3 \sigma$ samples (Obviously, the $3\sigma$ samples
are a subset of the surviving samples).
For each surviving sample, we also calculate the tuning extent
defined by $\Delta = Max\{|\partial \ln m_Z / \partial \ln p_i^{SUSY}|\}$ \cite{ft}, where $p_i^{SUSY}$ denotes
a soft breaking parameter at SUSY scale (fixed at $2$ TeV in this work).

For the convenience of our analysis, we categorize the surviving samples
into three cases according to the nature of the light Higgs scalar (note that
a doublet-dominated $h_1$ is ruled out by the LEP search for Higgs bosons and
$B \to X_s \gamma$):
\begin{itemize}
\item {\bf Case A}: The light scalar is the CP-odd $A_1$ ($A_1<h/2$) and it is singlet dominated.
\item {\bf Case B}: The light scalar is the CP-odd  $A_1$ ($A_1<h/2$) and it is doublet dominated.
\item {\bf Case C}: The light scalar is the CP-even $h_1$ ($h_1<h/2$) and it is singlet dominated.
\end{itemize}

\begin{table}
\caption{The favored parameter ranges for Case A, B and C in the NMSSM. In each item, the range in the
first row is for all surviving samples, and the second row corresponds to the $3\sigma$ samples (the null
result means the $3\sigma$ samples do not exist).} \centering
\begin{tabular}{|c|c|c|c|c|c|}
\cline{1-6}

\multicolumn{1}{|c|}{\multirow{2}{*}{}}
& \multicolumn{2}{c|}{Case~A  }
& \multicolumn{2}{c|}{Case~B  }
& Case~C \\
\cline{2-6}
\multicolumn{1}{|c|}{}
&~~$h_1$ is SM-like~ &~~$h_2$  is SM-like~
&~~$h_1$ is SM-like~ &~~$h_2$ is SM-like~
&~~$h_2$ is SM-like \\

\cline{1-6}
\multicolumn{1}{|c|}{\multirow{2}{*}{~$\lambda$~}}
& $0.1\thicksim0.75$ & $0.23\thicksim0.76$
& $0.1\thicksim0.25$ & $0.1\thicksim0.47$
& $0.20\thicksim0.74$ \\
\multicolumn{1}{|c|}{}
& $0.1\thicksim0.35$ & $0.23\thicksim0.72$
& --- & ---
& $0.22\thicksim0.74$ \\

\cline{1-6}
\multicolumn{1}{|c|}{\multirow{2}{*}{~~~$\kappa$~~~}}
& $ 0.1\thicksim0.65 $ & $ 0.1\thicksim0.25 $
& $ 0.1\thicksim0.54 $ & $ 0.32\thicksim0.6 $
& $ 0.1\thicksim0.46 $ \\
\multicolumn{1}{|c|}{}
& $ 0.1\thicksim0.63 $ & $ 0.1\thicksim0.23 $
& --- & ---
& $ 0.1\thicksim0.35 $ \\

\cline{1-6}
\multicolumn{1}{|c|}{\multirow{2}{*}{~~~$\tan\beta$~~~}}
& $ 1.4\thicksim30 $ & $ 1.6\thicksim15 $
& $ 6.5\thicksim12 $ & $ 2.8\thicksim7 $
& $ 1.7\thicksim18 $ \\
\multicolumn{1}{|c|}{}
& $ 5.2\thicksim30 $ & $ 4.2\thicksim15 $
& --- & ---
& $ 2.8\thicksim16 $ \\

\cline{1-6}
\multicolumn{1}{|c|}{\multirow{2}{*}{~~~$\mu$(GeV)}}
& $ 170\thicksim1000 $ & $ 108\thicksim270 $
& $ 390\thicksim1000 $ & $ 700\thicksim1000 $
& $ 110\thicksim450 $ \\
\multicolumn{1}{|c|}{}
& $ 198\thicksim610 $ & $ 115\thicksim235 $
& --- & ---
& $ 110\thicksim262 $ \\

\cline{1-6}
\multicolumn{1}{|c|}{\multirow{2}{*}{~~~$M_A$(GeV)}}
& $ 415\thicksim2000 $ & $ 310\thicksim2000 $
& $ 200\thicksim530 $ & $ 180\thicksim500 $
& $ 370\thicksim2000 $ \\
\multicolumn{1}{|c|}{}
& $ 850\thicksim2000 $ & $ 580\thicksim2000 $
& --- & ---
& $ 510\thicksim2000 $ \\

\cline{1-6}
\multicolumn{1}{|c|}{\multirow{2}{*}{~~~$M_P$(GeV)}}
& $ 1.3\thicksim160 $ & $ 37\thicksim135 $
& $ 220\thicksim550 $ & $ 1500\thicksim2000 $
& $ 110\thicksim475 $ \\
\multicolumn{1}{|c|}{}
& $ 10\thicksim80 $ & $ 40\thicksim130 $
& --- & ---
& $ 110\thicksim340 $ \\

\cline{1-6}
\multicolumn{1}{|c|}{\multirow{2}{*}{$M_2$(GeV)}}
& $ 100\thicksim670 $ & $ 290\thicksim1000 $
& $ 100\thicksim700 $ & $ 100\thicksim560 $
& $ 110\thicksim985 $ \\
\multicolumn{1}{|c|}{}
& $ 110\thicksim560 $ & $ 320\thicksim1000 $
& --- & ---
& $ 160\thicksim965 $ \\

\cline{1-6}
\multicolumn{1}{|c|}{\multirow{2}{*}{$M_{Q_3}$(GeV)}}
& $ 205\thicksim2000 $ & $ 215\thicksim2000 $
& $ 280\thicksim2000 $ & $ 100\thicksim1500 $
& $ 505\thicksim2000 $ \\
\multicolumn{1}{|c|}{}
& $ 345\thicksim2000 $ & $ 330\thicksim2000 $
& --- & ---
& $ 585\thicksim1980 $ \\

\cline{1-6}
\multicolumn{1}{|c|}{\multirow{2}{*}{$M_{U_3}$(GeV)}}
& $ 180\thicksim2000 $ & $ 400\thicksim2000 $
& $ 200\thicksim2000 $ & $ 100\thicksim2000 $
& $ 500\thicksim2000 $ \\
\multicolumn{1}{|c|}{}
& $ 235\thicksim2000 $ & $ 400\thicksim 2000 $
& --- & ---
& $ 570\thicksim2000 $ \\

\cline{1-6}
\multicolumn{1}{|c|}{\multirow{2}{*}{$A_{t}$(GeV)}}
& $ -4960\thicksim4920 $ & $ -5000\thicksim5000 $
& $ -5000\thicksim-2000 $ & $ -3000\thicksim-300 $
& $ -4960\thicksim4980 $ \\
\multicolumn{1}{|c|}{}
& $ -4400\thicksim4630 $ & $ -5000\thicksim5000 $
& --- & ---
& $ -4960\thicksim4870 $ \\

\cline{1-6}
\multicolumn{1}{|c|}{\multirow{2}{*}{$M_{\tilde{l}}$(GeV)}}
& $ 100\thicksim1000 $ & $ 100\thicksim500 $
& $ 100\thicksim750 $ & $ 100\thicksim350 $
& $ 100\thicksim800 $ \\
\multicolumn{1}{|c|}{}
& $ 100\thicksim1000 $ & $ 100\thicksim500 $
& --- & ---
& $ 100\thicksim800 $ \\

\cline{1-6}
\multicolumn{1}{|c|}{\multirow{2}{*}{~~~$A_\lambda$(GeV)}}
& $ -2500\thicksim1920 $ & $ -550\thicksim2180 $
& $ -2000\thicksim-400 $ & $ -3700\thicksim-800 $
& $ 300\thicksim2150 $ \\
\multicolumn{1}{|c|}{}
& $ -600\thicksim820 $ & $ -550\thicksim2100 $
& --- & ---
& $ 465\thicksim2000 $ \\

\cline{1-6}
\multicolumn{1}{|c|}{\multirow{2}{*}{~~~$A_\kappa$(GeV)}}
& $ -34\thicksim75 $ & $ -38\thicksim61 $
& $ -70\thicksim-16 $ & $ -1300\thicksim-200 $
& $ -620\thicksim-70 $ \\
\multicolumn{1}{|c|}{}
& $ -5\thicksim0.16 $ & $ -35\thicksim35 $
& --- & ---
& $ -395\thicksim-70 $ \\

\cline{1-6}
\end{tabular}
\end{table}

\subsection{Case A ($A_1<h/2$, singlet dominated)}

\begin{figure}[]
\includegraphics[width=12.0cm]{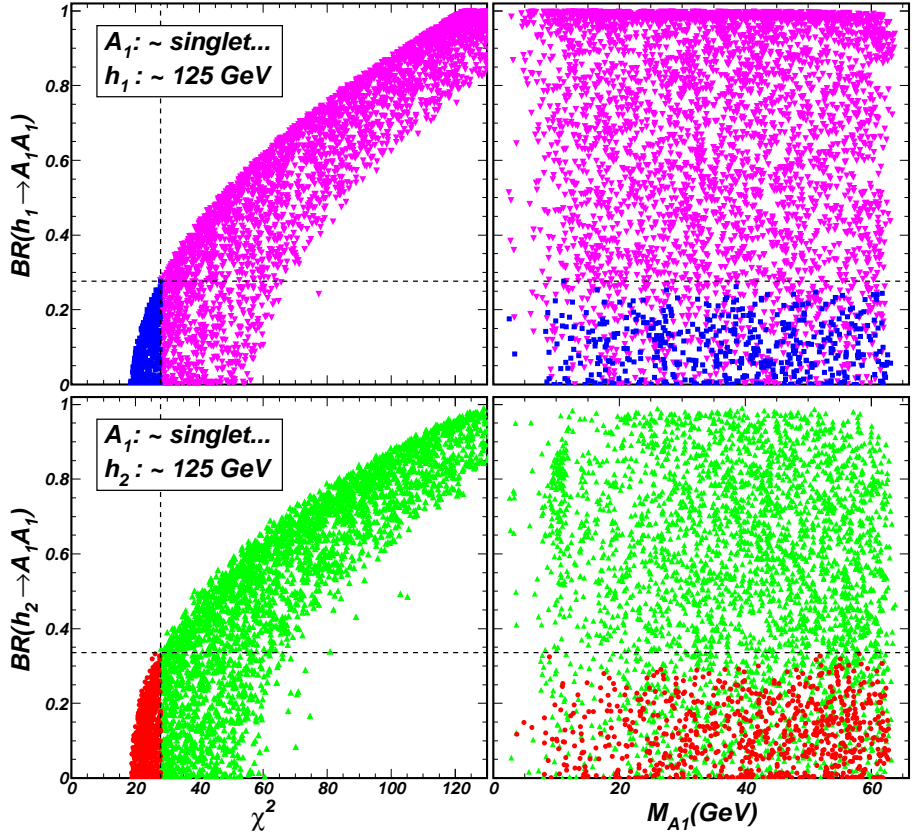}\vspace{0.1cm}
\vspace*{-0.5cm}
\caption{The scatter plots of the surviving samples in Case A projected on the plane of $\chi^2$ versus
$Br(h \to A_1 A_1)$ ($h$ denotes the SM-like Higgs boson) and the plane of $m_{A_1}$ versus
$Br(h \to A_1 A_1)$ respectively. The upper panel is for the `SM-like $h_1$' scenario
with the $3\sigma$ samples marked out as squares (blue),
and the bottom panel is for the  the `SM-like $h_2$' scenario with the $3\sigma$ samples
marked out as circles (red).}
\label{fig1}
\end{figure}

\begin{figure}[]
\includegraphics[width=12.0cm]{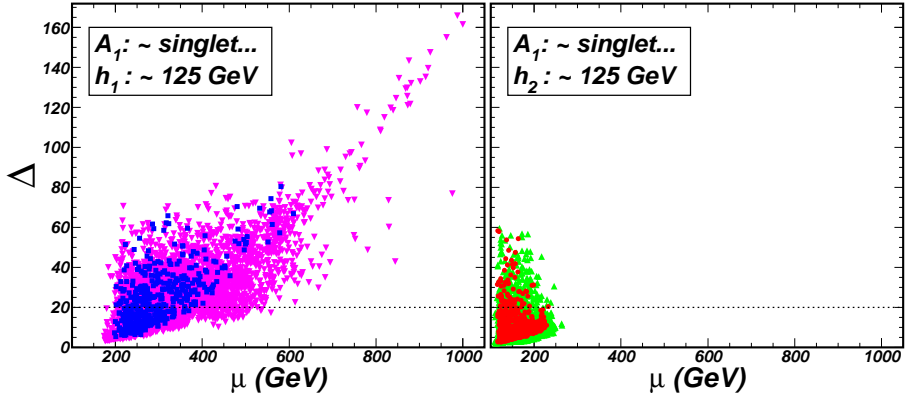}\vspace{0.1cm}
\vspace*{-0.5cm}
\caption{Same as Fig.\ref{fig1}, but projected on the $\mu$ versus $\Delta$ plane.}
\label{fig2}
\end{figure}

In Case A, the SM-like 125 GeV Higgs boson $h$ may be either the lightest CP-even state $h_1$
or the next-to-lightest CP-even state $h_2$. In Table I, we list the favored parameter ranges for
all the surviving samples and the $3\sigma$ samples in Case A. 
We note that in this case the parameter $\tan \beta$ can be very large \cite{NMSSM-large-tanb}. 
This table indicates that in each scenario
the ranges of some parameters for the surviving samples are significantly
wider than the corresponding $3\sigma$ samples.
Furthermore, we compare the number of all the surviving samples with the $3\sigma$ samples,
and find that the latter is at most one fifth of the former.
These facts reflect that the current LHC Higgs data can severely constrain
the parameter space of the NMSSM.
This table also indicates that, in order to predict a light singlet-dominated
$A_1$, the value of $M_P$ should be less than 160 GeV.

From analyzing the surviving samples, we find two features for Case A:
\begin{itemize}
\item One is that the $\chi^2$ value in the fit of the Higgs data
may be rather low with $\chi^2_{min}\simeq 17$ for $24$ sets of experimental data,
and it increases as the branching ratio of the exotic decay $ h \to A_1 A_1$ becomes larger.
This feature is exhibited in Fig.\ref{fig1}.
This figure reflects the fact that the NMSSM can explain the Higgs data quite well
given that $Br(h\to A_1 A_1)$ is moderately small. This figure also reveals the information
that, without the Higgs data, $Br(h\to A_1 A_1)$ can exceed $90\%$,
while after considering the constraints from the Higgs data at $3\sigma$,
it is less than $28\%$ for $h_1$ being the SM-like Higgs (the `SM-like $h_1$' scenario) and $34\%$ for $h_2$ being the SM-like Higgs (the `SM-like $h_2$' scenario).
This conclusion is independent of the value of $m_{A_1}$.
As a comparison, we checked that for any exotic decays of the Higgs boson
(with the SM Higgs couplings to fermions and gauge bosons), the Higgs data
restrain the exotic decay branching ratio to be less than $28\%$ at $3\sigma$ level.
This result can be seen as an update of that in \cite{1302-Belanger} after the Rencontres de Moriond 2013, 
but different from those in \cite{1306-Ananthanarayan} for different data treatments. 

\item The other feature is that the
tuning extent $\Delta$ can be less than 10, reflecting that the NMSSM is quite natural.
This feature is shown in Fig.\ref{fig2}.
Compared with the `SM-like $h_1$' scenario,
a lower $\Delta$ is predicted for the `SM-like $h_2$' scenario.
This is because $m_Z$ is sensitive to the
value of $\mu$ (note the tree level relation
$ m^2_{Z}= 2 ( m^2_{H_d} - m^2_{H_u} \tan^{2}\beta)/(\tan^{2}\beta-1 ) - 2 \mu^{2} $
with $m^2_{H_d}$ and $m^2_{H_u}$ representing the weak scale soft SUSY breaking masses
of the Higgs fields \cite{ft}), and for the `SM-like $h_2$' scenario a lower $\mu$
is preferred.
\end{itemize}

\begin{figure}[]
\includegraphics[width=8.0cm]{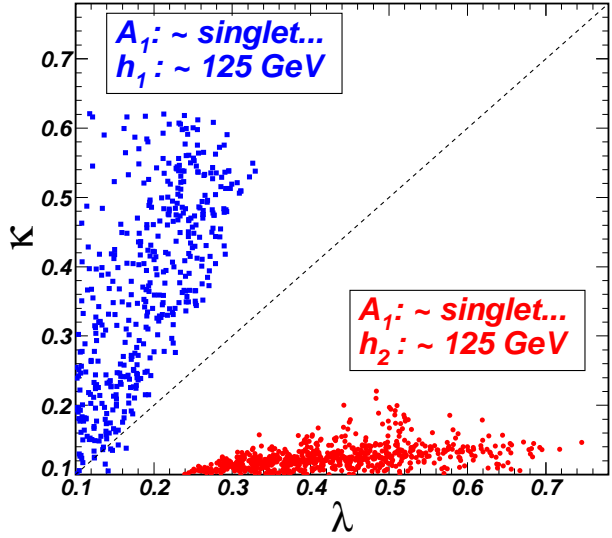}\vspace{0.1cm}
\vspace*{-0.5cm}
\caption{Same as Fig.\ref{fig1}, but projected on the $\lambda$ versus $\kappa$ plane
(here only the $3\sigma$ samples are ploted).}
\label{fig3}
\end{figure}
\begin{figure}[]
\includegraphics[width=8.0cm]{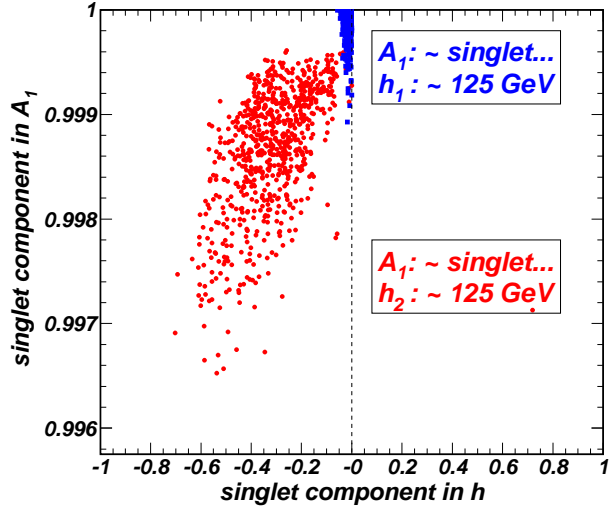}\vspace{0.1cm}
\vspace*{-0.5cm}
\caption{Same as Fig.\ref{fig3}, but show the singlet component coefficients of $A_1$ and $h$.}
\label{fig4}
\end{figure}
\begin{figure}[]
\includegraphics[width=12.0cm]{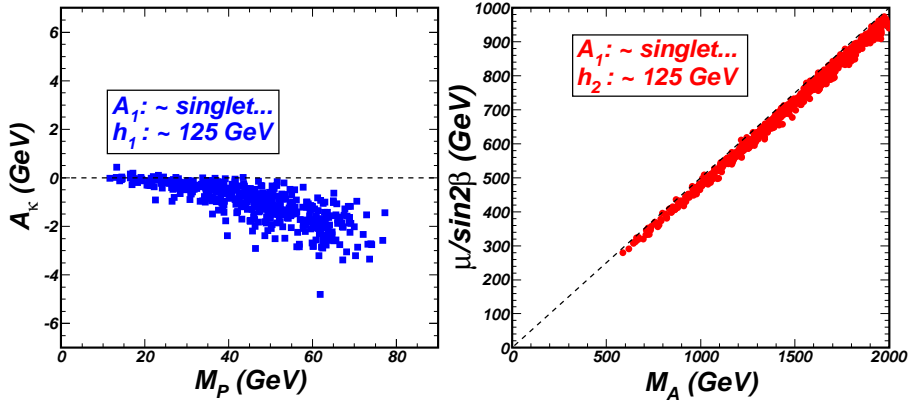}\vspace{0.1cm}
\vspace*{-0.5cm}
\caption{Same as Fig.\ref{fig3}, but showing the correlation of $M_P$ with $A_\kappa$
for  the `SM-like $h_1$' scenario (left panel)
and the correlation of $M_A$ with $\mu/\sin 2\beta$ for the `SM-like $h_2$' scenario
(right panel).}
\label{fig5}
\end{figure}

About Case A, more points should be noted. (i) The first is that
the `SM-like $h_1$' scenario and the `SM-like $h_2$' scenario actually correspond to two distinct parameter
regions of the NMSSM. To illustrate this point, we consider the parameters $\lambda$ and $\kappa$
and project the $3\sigma$ samples on the $\lambda$ versus $\kappa$ plane in Fig.\ref{fig3}.
This figure indicates that, in contrast with the fact that most samples for the `SM-like $h_1$' scenario
satisfy $\lambda \lesssim \kappa$, the `SM-like $h_2$' scenario is characterized by
$\lambda \gg \kappa$. The reason is that as far as the $3\sigma$ samples are concerned,
$M_{33}$ in Eq.(\ref{MS}) is approximated by $M_{33} \simeq 4 (\kappa v_s)^2 = 4(\kappa \mu/\lambda)^2$.
Given $\mu > 100 {\rm GeV}$ as required by the LEP bound on chargino mass, $\lambda$ should be much
larger than $\kappa$ to guarantee $M_{22}^2 > M_{33}^2$, which is a necessary condition to predict
$h_2 \sim 125 ~{\rm GeV}$. (ii) The second point is that $A_1$ should be highly singlet dominated
and the properties of the SM-like Higgs boson for the `SM-like $h_1$' scenario and
the `SM-like $h_2$' scenario may be quite different. To exhibit this conclusion, we show in Fig.\ref{fig4}
the singlet component coefficients of $A_1$ and $h$ for the $3\sigma$ samples.
This figure indicates that the singlet component coefficient of $A_1$ (i.e. $U_{12}$)
is larger than 0.99 for both scenarios. This figure also indicates that the SM-like $h_1$
has a very small singlet component (i.e. $V_{13} \sim 1\%$) while
the SM-like $h_2$ may have a sizable singlet component with the
corresponding coefficient $V_{23}$ reaching 0.7.
In fact, we checked that the $hb\bar{b}$ coupling is
approximately equal to the SM value for the `SM-like $h_1$' scenario
 and may be much smaller for the `SM-like $h_2$' scenario.
About Case A, we remind that, due to the singlet nature of $A_1$, the $hA_1A_1$ interaction should be
very weak, but on the other hand, since the total width of the SM-like Higgs boson is also small
(about $4~{\rm MeV}$ in the SM), $Br(h \to A_1 A_1)$ may still be sizable.
(iii) The last point is that, since we require the theory to predict a light scalar
and meanwhile satisfy various experimental constraints,
some parameters are limited in certain narrow ranges or correlate with other parameters, as shown
in Fig.\ref{fig5}.
The left panel indicates that in the `SM-like $h_1$' scenario we have $A_\kappa \simeq 0$,
and the right panel shows that in the `SM-like $h_2$' scenario we have $M_A \sin 2\beta/\mu \simeq 2$.
We checked that a very small $A_\kappa$
is needed to predict a light singlet dominated $A_1$, while the correlation $M_A \sin 2\beta/\mu \simeq 2$
is characteristic in predicting $h_2 \simeq 125~{\rm GeV}$, as observed in \cite{125-cao}.

\subsection{Case B ($A_1<h/2$, doublet dominated)}

\begin{figure}[]
\includegraphics[width=12.0cm]{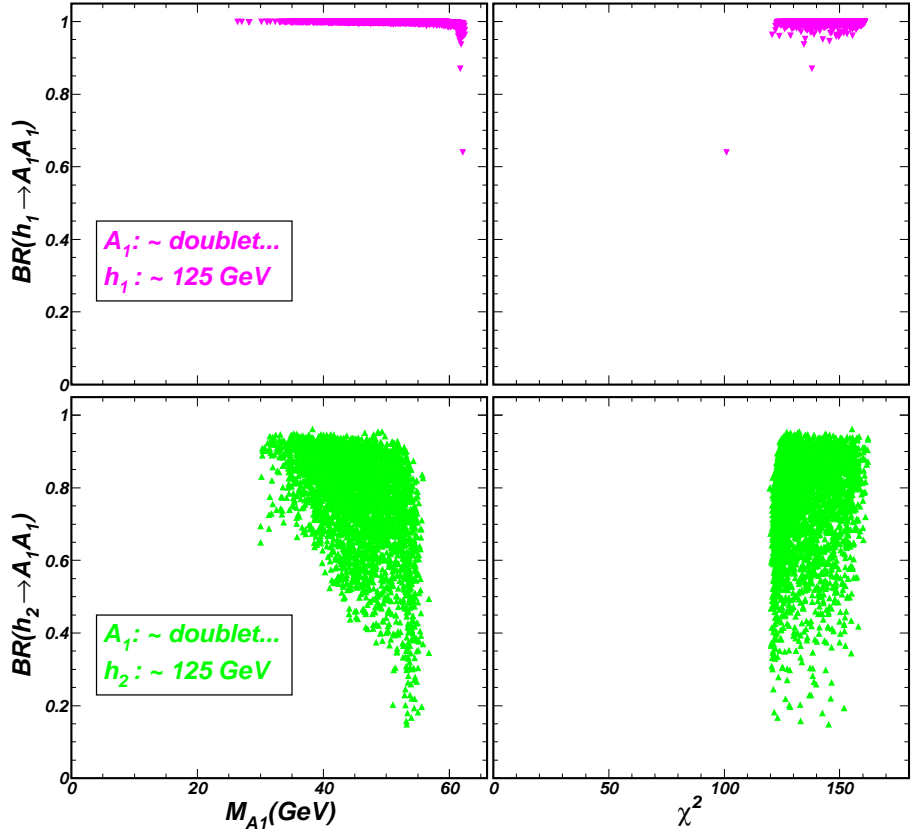}\vspace{0.1cm}
\vspace*{-0.5cm}
\caption{Same as Fig.\ref{fig1}, but showing the surviving samples in Case B (no $3\sigma$ samples).}
\label{fig6}
\end{figure}

\begin{figure}[]
\includegraphics[width=12.0cm]{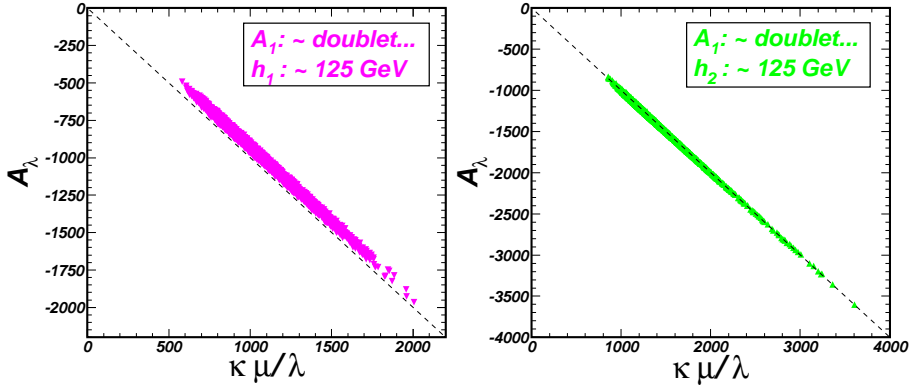}\vspace{0.1cm}
\vspace*{-0.5cm}
\caption{Same as Fig.\ref{fig5}, but showing the correlation of $\kappa \mu/\lambda$ with $A_\lambda$ for the surviving samples in Case B.}
\label{fig7}
\end{figure}

As in Case A, the SM-like 125 GeV Higgs boson in Case B may be either $h_1$ or $h_2$,
and the corresponding favored parameter regions of the surviving samples are shown in Table I.
We emphasize that the parameter $M_A$ in this table is defined at the scale of 2 TeV,
and in calculating the CP-odd Higgs boson masses by the NMSSMTools we use the value at
the mass scale of the third generation squarks which can be obtained by the
renormalization group equation.
Moreover, we checked that the surviving samples are characterized by a relatively large matrix
element $M_{-12}^2$ in Eq.(\ref{MA}). This is helpful to suppress the mass of $A_1$.

In Fig.\ref{fig6} we project the surviving samples on the plane of $M_{A_1}$ versus $Br(h \to A_1 A_1)$
and the plane of $\chi^2$ versus $Br(h \to A_1 A_1)$ respectively.
This figure indicates that in the `SM-like $h_1$' scenario, the branching ratio
of the decay $h \to A_1A_1$ is always larger than $60\%$ so that $\chi^2 > 100$,
while in the  `SM-like $h_2$' scenario, although the rate of
the decay $h \to A_1A_1$ may be small, e.g. about $10\%$ for $m_{A_1} \simeq 55~{\rm GeV}$,
the $\chi^2$ value is still larger than 100.
The reason is that the $h b \bar{b}$ coupling in the `SM-like $h_2$' scenario
is at least one times larger than its SM prediction.
In fact, the `SM-like $h_2$' scenario in Case B actually corresponds to a non-decoupling region
of the NMSSM since the mass of the charged Higgs boson varies from $130~{\rm GeV}$ to $150~{\rm GeV}$.
Consequently, the properties of the SM-like Higgs boson are expected to deviate greatly from
the SM prediction. To summarize, Fig.\ref{fig6} indicates that Case B is actually
disfavored by the fit of the Higg data (no $3\sigma$ samples exist).

Also as in Case A, a strong correlation between some parameters is needed to predict a doublet
dominated $A_1$. In Fig.\ref{fig7} we show the correlation between the parameter $A_\lambda$ and
the parameter $\kappa \mu/\lambda$  for the surviving samples in this case.
From Eq.(\ref{MA}), one can infer that such a correlation is needed to reduce the value of $M_A$.

\subsection{Case C  ($h_1<h/2$, singlet dominated)}

\begin{figure}[]
\includegraphics[width=12.0cm]{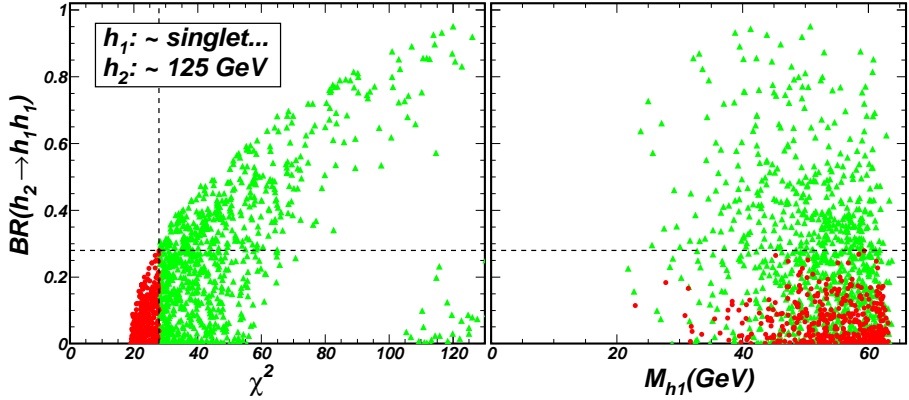}\vspace{0.1cm}
\vspace{0.1cm}
\caption{Same as Fig.\ref{fig1}, but for Case C where $h_2$ is the SM-like Higgs boson.}
\label{fig8}
\end{figure}

\begin{figure}[]
\includegraphics[width=8.0cm]{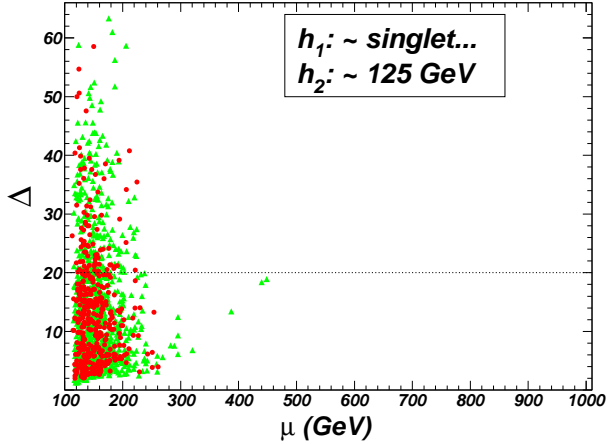}\vspace{0.1cm}
\vspace*{-0.5cm}
\caption{Same as Fig.\ref{fig2}, but for Case C.}
\label{fig9}
\end{figure}

\begin{figure}[]
\includegraphics[width=8.0cm]{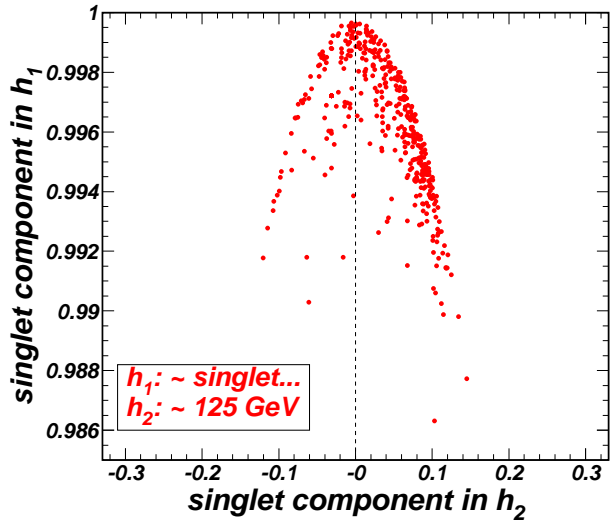}\vspace{0.1cm}
\vspace*{-0.5cm}
\caption{Same as Fig.\ref{fig4}, but for the singlet component coefficients for $h_1$ and $h_2$ in Case C.}
\label{fig10}
\end{figure}

In Case C the SM-like Higgs boson is the next-to-lightest CP-even state $h_2$,
and due to the strong constraints from the LEP search for Higgs bosons and $B \to X_s \gamma$,
a doublet dominated $h_1$ is actually ruled out.
In Table I, we show the favored parameter regions for the surviving samples and
also the $3\sigma$ samples.
As pointed out in \cite{NMSSM-lightDM}, in order to predict a light $h_1$,
one only needs to tune the value of $A_\kappa$ when other parameters are fixed.
So, except for the correlation shown on the left panel of Fig.\ref{fig5} and the condition
$\kappa < \lambda$ which is necessary to predict $m_{h_2} \simeq 125 ~{\rm GeV}$,
there is no other special features for the parameters of Case C.

Like the `SM-like $h_2$' scenario in Case A, the $\chi^2$ value and the parameter $\Delta$ may be
as low as about 17 and 10 respectively.
These features are presented in Fig.\ref{fig8} and Fig.\ref{fig9}.
About Case C, one should note that the branching ratio of $h \to h_1 h_1$ should be less than $28\%$
at $3 \sigma$ level (see Fig.\ref{fig8}). One should also note that, as shown in Fig.\ref{fig10}
where the singlet component coefficients of $h_1$ and $h_2$ are presented for the $3\sigma$ samples,
$h_1$ in Case C is highly singlet dominated while $h_2$ is highly doublet dominated.

In summary, one may conclude that the current experiments still allow for
the existence of a light scalar (CP-even or CP-odd).  But the LHC Higgs data
have required it to be highly singlet dominated.
Moreover, in the NMSSM either $h_1$ or $h_2$ may play the role of the SM-like Higgs boson $h$,
and for each case the properties of $h$ may be quite different.

\section{Detection of a light scalar at future colliders}
As discussed in the preceding section, if there exists a light scalar with mass lighter than half
the SM-like Higgs boson mass in the NMSSM, it should be highly singlet dominated.
Consequently, its interactions with the fermions and the gauge bosons in the SM
are very weak, which implies that this scalar is difficult to search at colliders.
But on the other hand, although the interaction of this scalar with the SM-like Higgs boson
is also weak, the rate of $h$ decay into the scalar pair may still be sizable due to the narrow
width of $h$.
This fact motivates us to scrutinize the decay product of $h$ to search for the light scalar.
In the following, we take Case A as an example to discuss the prospect of such a search
via different processes at colliders.

\begin{figure}[]
\includegraphics[width=12.0cm]{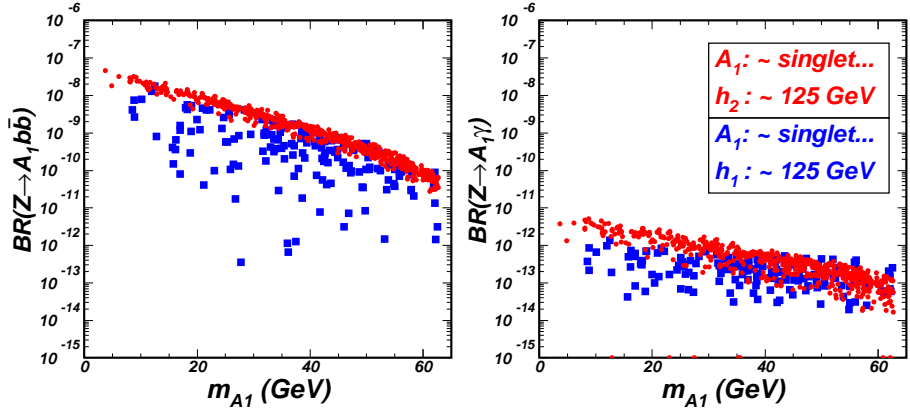}\vspace{0.1cm}
\vspace*{-0.5cm}
\caption{The branching ratios of the rare Z decays $Z \to A_1 b \bar{b}$ and $Z \to A_1 \gamma$
as a function of $m_{A_1}$ for the $3\sigma$ samples in Case A. The squares (blue) correspond to
the results of the `SM-like $h_1$' scenario, and the bullets (red) are for the `SM-like $h_2$' scenario.}
\label{fig11}
\end{figure}

\begin{figure}[]
\includegraphics[width=8.0cm]{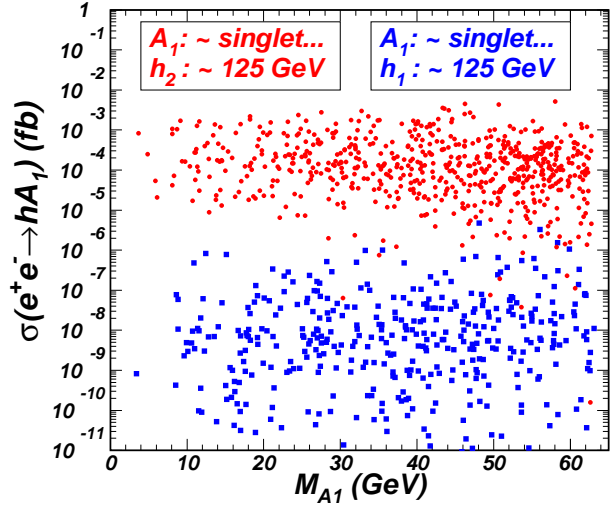}\vspace{0.1cm}
\vspace*{-0.5cm}
\caption{Same as Fig.\ref{fig11}, but for the cross section of $hA_1$ associated production at an
electron-positron collider with $\sqrt{s} = 250~{\rm GeV}$.}
\label{fig12}
\end{figure}

First, we consider the light $A_1$ comes from the $Z$-decay.
For this end, we calculate the branching ratios of the rare decays
$Z \to A_1 b \bar{b}$ and $Z \to A_1 \gamma$ with the code of our previous work \cite{RareZdecay}
and show these ratios in Fig.\ref{fig11}.
This figure indicates that, as far as the $3\sigma$ samples in Case A are concerned,
the ratios are at most $10^{-8}$ and $10^{-12}$, respectively.
Since the dominant decay product of $A_1$ is $b \bar{b}$ with a branching ratio
being about $90\%$, the main signals of the decays are $b\bar{b} b \bar{b}$
and $b\bar{b}\gamma$, respectively. Then, compared with the LEP uncertainties on these signals,
we learn that the ratios are at least $10^{-4}$ lower than the LEP sensitivity \cite{PDG2012}.

Second, we consider the $hA_1$ associated production at an electron-positron collider
with $\sqrt{s}=250~{\rm GeV}$.
In Fig.\ref{fig12} we show the production rate as a function of $m_{A_1}$.
Obviously, since the rate is maximally at the order of $10^{-3}$ fb,
this associated production process can hardly be utilized to search for the scalar.

Next we investigate the possibility of searching for $A_1$
at the LHC via the decay $h\to A_1 A_1 \to 4 b$.
Such an issue has been discussed in \cite{Cheung:2007sva,Tao-Han} and it was found
that the process $pp \to V h \to l + 4 b + X$ ($V=W,Z$, $l$ denotes one lepton and $X$ denotes anything)
is well suited for such a search. In this work, we fix $m_h = 125~{\rm GeV}$ and perform an analysis
as in \cite{Cheung:2007sva}.
The signal contains at least one isolated lepton, $e$ or $\mu$, and exactly 4 $b$-tagged jets.
The corresponding backgrounds mainly come from the $t\bar{t}$ production  with one top quark
decaying hadronically and the other top quark decaying semi-leptonically, the $t\bar{t}b\bar{b}$
production  with some of the top quark decay products missed, the $t\bar{t}c\bar{c}$ production
with the charm quark jets mistagged as bottom quark jets and also the $W/Z + 4 b$  production
processes. In our simulation, the signal and background processes are modeled with
MadGraph 5 \cite{Alwall:2011uj}, which incudes Pythia 6.4 \cite{Sjostrand:2006za} for initial
and final state radiation, parton shower and hadronization, and pass through the fast detector
simulation with DELPHES \cite{deFavereau:2013fsa}.
Jets are reconstructed with FastJet \cite{Cacciari:2011ma,hep-ph/0512210} by using the anti-$k_T$
algorithm with a distance parameter of 0.5. The cuts we considered are:
\begin{itemize}
\item The basic cuts:
\begin{eqnarray}
&&p_T(j) \geq 15~{\rm GeV},  \quad |\eta(j)| \leq 2.5, \quad  p_T(l) \geq 15~{\rm GeV}, \quad |\eta(l)| \leq 2.5, \\ && \Delta R(b,b) \geq 0.4, \quad \Delta R(b,l) \geq 0.4, \nonumber
\end{eqnarray}
where $p_T$ denotes the transverse momentum, $\eta$ represents pseudorapidity and
$\Delta R (b,j) =\sqrt{(\Delta \eta)^2+(\Delta \phi)^2}$ is the angular separation
of the b-jet and the particle $j$ ($j=b,l$).
\item $|M_{4b} -115|\leq 15~{\rm GeV}$ with $M_{4b}$ denoting the invariant mass of the four bottom
quarks. This cut is motivated by the fact that the four bottom quarks originate from the SM-like
Higgs boson decay, and due to possible momentum missing in the jet reconstruction, $M_{4b}$ is
peaked at about $115~{\rm GeV}$ instead of at the Higgs boson mass \cite{Momentum-miss}.
\end{itemize}
Moreover, in order to get a realistic estimation of the signal and backgrounds,
we also assume a b-tagging efficiency of $70\%$ for
a bottom quark jet and a mis-tagging probability of $5\%$ ($1\%$)
for a charm quark jet (light quark or gluon jet).

\begin{figure}[tH]
\centering
\includegraphics[width=12cm]{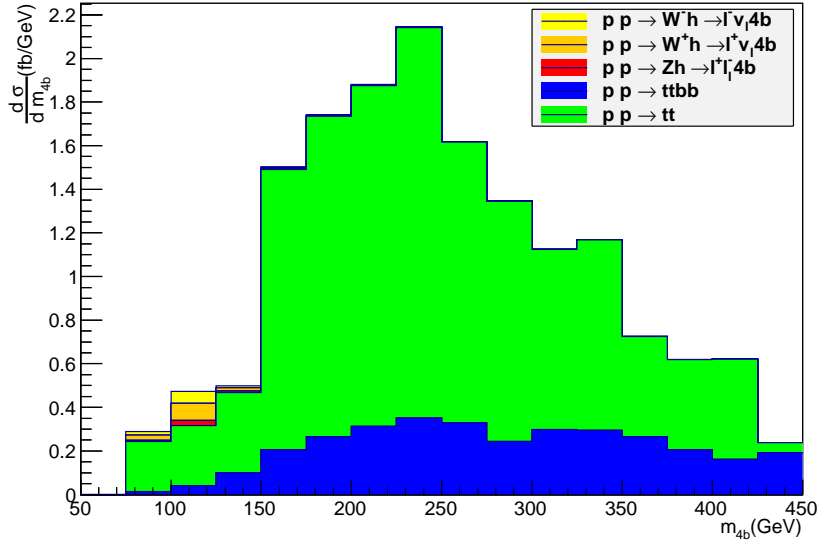}
\vspace{-0.5cm}
\caption{The invariant mass $M_{4b}$ distribution for the signal and the backgrounds
after the basic cuts. Here we fix $m_{A_1} = 45~{\rm GeV}$ and $C_{4b}^2=0.33$.}
\label{fig13}
\end{figure}

\begin{table*}[]
\centering
\caption{\small The rates of the signal and various backgrounds after different cuts
for $m_{A_1} = 45~{\rm GeV}$ and $C_{4b}^2=0.33$.}
\medskip
\begin{ruledtabular}
\begin{tabular}{llllllllll}
                                 & $t\bar{t}$ & $t\bar{t}+b\bar{b}$ & $t\bar{t}+c\bar{c}$ & V+jets & Total bkg & Zh    & $W^+h$ & $W^-h$ & Total signal\\
\hline
  $\sigma_{basic~cuts} (fb$)    &  12.45     & 3.28                  &  0.039               & 0.264  & 16.13     & 0.049 & 0.133  & 0.087  &  0.26  \\
  $\sigma_{M_{4b}~cut} (fb$)    &  0.170     & 0.031                 &  0.00045             & 0.016  & 0.22      & 0.024 & 0.095  & 0.053  &  0.17 \\
\end{tabular}
\end{ruledtabular}
\end{table*}

Noticing that the signal rate after the cuts depends only on an overall scaling factor
\begin{equation}
C_{4b}^{2}=(\frac{g_{VVh}^{NMSSM}}{g_{VVh}^{SM}})^2 \times Br(h\rightarrow A_1 A_1) \times (Br(A_1\rightarrow b\bar{b}))^2,
\label{c4b}
\end{equation}
which determines the cross section of the process  $pp \to V h \to V 4 b$ at the LHC,
and the mass of $A_1$ which determines the cut efficiency,
we fix $m_{A_1} = 45~{\rm GeV}$ and $C_{4b}^2=0.33$, and illustrate the distributions of $M_{4b}$
for both the signal and various backgrounds in Fig.\ref{fig13}.
We also list the rates of the signal and the backgrounds after different cuts in Table II.
These results indicate that the $M_{4b}$ cut is very efficient in suppressing the backgrounds,
and also that the $t\bar{t}$ background is still dominant over other backgrounds after the cut.
Moreover, for the benchmark point we considered,
we estimate that its significance $S/\sqrt{B}$ is about 6.37
for an integrated luminosity of $300 ~{\rm fb^{-1}}$.

\begin{figure}[tH]
\includegraphics[width=10cm]{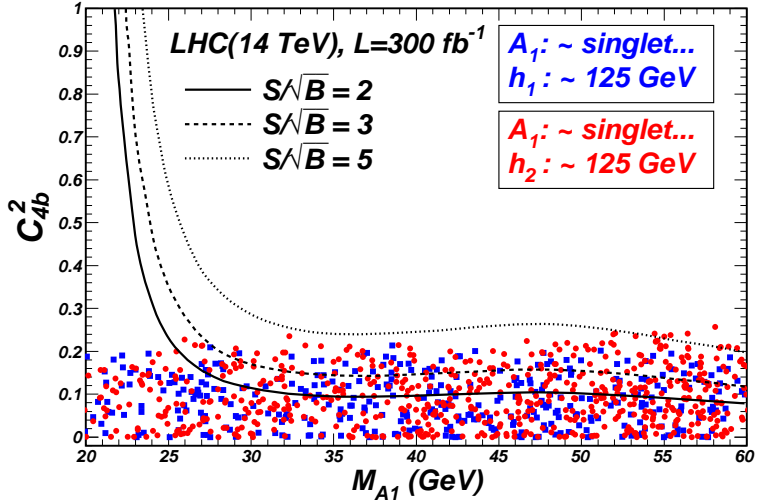}
\vspace{-0.5cm}
\caption{Same as Fig.11, but projected on the plane of $M_{A_1}$ versus $C^2_{4b}$. The significance of the LHC-14 for $300 ~{\rm fb^{-1}}$ integrated luminosity is also plotted on this plane.
}
\label{fig14}
\end{figure}

In order to exhibit the capability of the LHC in the $A_1$ search, in Fig.\ref{fig14} we plot the $3\sigma$ samples together with the significance curves of
$S/\sqrt{B}=2,3,5$ for an luminosity of $300~{\rm fb^{-1}}$ on the $m_{A_1}$ versus $C_{4b}^2$ plane.
This figure shows that in order to
discover the light scalar, $C_{4b}^2$ should be larger than 1
for $m_{A_1} \lesssim 25~{\rm GeV}$, and with the increase
of $m_{A_1}$, the requirement on $C_{4b}^2$ decreases to 0.2 for $ m_{A_1} = 60~{\rm GeV}$.
We can also see that nearly all of the $3\sigma$ samples in the two scenarios are under the $S/\sqrt{B}=5$ curve,
which means that in order to discover the light scalar
a luminosity over $300~{\rm fb^{-1}}$ is needed.

Compared with the simulation result in \cite{Cheung:2007sva}, we note our significance is much lower.
The reason is that the authors of \cite{Cheung:2007sva} performed the simulation at parton level,
while in our analyse we considered the initial and final state radiation, the parton shower and
the hadronization effect with Pythia, the detector effect with DELPHES,
and the reconstruction of jets with FastJet.
Consequently, the $M_{4b}$ distribution of the $t\bar{t}$ production moves towards lower end
so that the $t\bar{t}$ production is still the dominant background after the cuts.
This is quite different from the results of \cite{Cheung:2007sva}
where the main background comes from the $t\bar{t}b\bar{b}$ production.
Another consequence of our treatment is that the jet reconstruction can hurt both the signal
and the backgrounds greatly, especially for our case where the signal contains exactly four b-jets.
We checked that if we perform the simulation at parton level as in \cite{Cheung:2007sva},
we can reproduce its results.

\begin{figure}[]
\includegraphics[width=12.0cm]{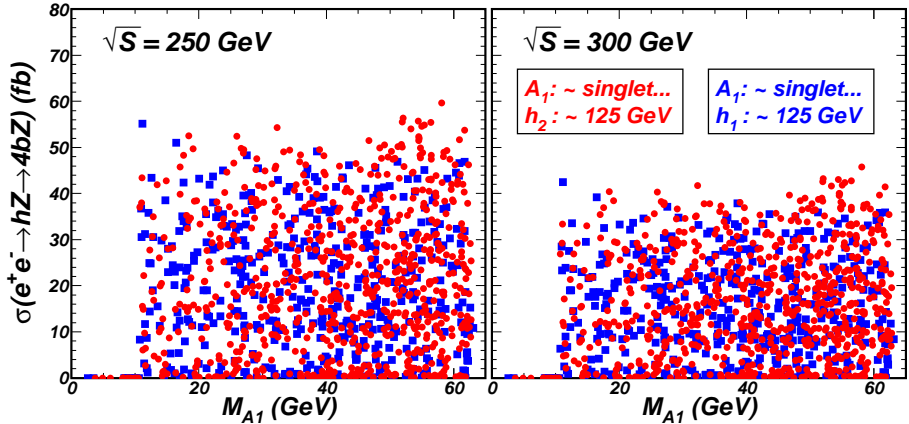}\vspace{0.1cm}
\vspace*{-0.5cm}
\caption{Same as Fig.\ref{fig12}, but for the cross section of the process $e^+ e^- \to Z h \to Z 4 b$ at an
electron-positron collider with $\sqrt{s} = 250~{\rm GeV}$ and $300~{\rm GeV}$ respectively.}
\label{fig15}
\end{figure}
Finally, since the properties of $h$ can be precisely measured though the $Zh$ associated production at an electron-positron collider,
we also calculate the cross section of the process $e^+e^- \to Z h \to Z 4b$ for a collision energy
$\sqrt{s}=250~{\rm GeV}$ and  $300~{\rm GeV}$ respectively. The results are shown in Fig.\ref{fig15}.
This figure indicates that, as far as the $3\sigma$ samples in Case A are concerned,
the rate can be as large as $56 ~{\rm fb}$ for $\sqrt{s}=250~{\rm GeV}$.
Compared with the same final state at the LHC with $\sqrt{s} = 14 ~{\rm TeV}$, although
such a production rate is only about one fourth, the signal is free of the backgrounds listed in Table II.
So a rather low prodcution rate at an electron-positron collider may result in the $A_1$ discovery.
We checked that a production rate over $10 ~{\rm fb}$ corresponds to $C_{4b}^2 > 0.04$ (such a small $C_{4b}^2$
is not accessible at the LHC for $300 ~{\rm fb^{-1}}$ integrated luminosity).
Fig.\ref{fig15} also indicates that, since the $Zh$
associated production is a $s$-channel process, the signal rate decreases as the increase of the
collision energy.

\section{Conclusion}

In the NMSSM, due to the introduction of one new gauge singlet Higgs field, one of the neutral Higgs scalars
(CP-even or CP-odd) may be lighter than half the SM-like Higgs boson. In this case, the SM-like Higgs boson
$h$ can decay into the scalar pair and consequently the visible $\gamma \gamma$ and $Z Z^\ast$ signal rates
at the LHC will be suppressed. In this work, we checked the constraints of the latest LHC Higgs data on such
a possibility. First, we scanned comprehensively the parameter space of the NMSSM by considering various
experimental constraints. Then we focused on the surviving samples which predict a light scalar.
According to the properties of the scalar, we categorized the samples into three cases calsses:
Case A ($A_1<h/2$, singlet dominated),  Case B  ($A_1<h/2$, doublet dominated) and Case C ($h_1<h/2$,
singlet dominated).
For the surviving samples we performed a fit using the latest LHC Higgs data. We found that
the Higgs data can severely constrain the parameter space, e.g., for Case A and Case C,
less than one fifth of the surviving samples are allowed by the Higgs data at $3\sigma$ level,
and for Case B all samples are actually ruled out.
We further focused on the $3\sigma$ samples allowed by the Higgs data and analysed the properties of the
light scalar, including its favored parameter region, its composition as well as the ratio of $h$ decay
into the scalar pair.  Finally, we examined the detection of such a scalar at future colliders.
From our analysis we obtained the following observations:
\begin{itemize}
\item[(i)] Without the LHC Higgs data, the light Higgs boson $A_1$ can be either singlet-dominated
or doublet-dominated; while after considering the constraints from the Higgs data, it should be highly singlet dominated.
\item[(ii)] In the `SM-like $h_1$' and `SM-like $h_2$' scenarios of Case A,
the Higgs data require the branching ratio of $h \to A_1 A_1$ to be less than
$28\%$ and $34\%$ respectively;  while in the `SM-like $h_2$' scenario of Case C,
the Higgs data require the ratio of $h \to h_1 h_1$ to be less than $28\%$.
\item[(iii)] An efficient way at the LHC to detect the light scalar is through the $Vh$ ($V=W,Z$)
associated production with $h$ decaying exotically into four bottom quarks.
A detailed Monte Carlo simulation indicates that, if the branching ratio
of the exotic decay is less than $30\%$, more than $300 ~{\rm fb^{-1}}$ luminosity is needed to discover the scalar.
At a future electron-positron collider with $\sqrt{s} \simeq 250~{\rm GeV}$,
the capability to detect the light scalar may be greatly improved
by looking for the process $e^+e^- \to Z h \to Z A_1 A_1 \to Z 4 b$.
\end{itemize}

\section*{Acknowledgement}
This work was supported in part by the National Natural
Science Foundation of China (NNSFC) under grant Nos. 10821504,
11135003, 10775039, 11075045, by Specialized Research Fund for
the Doctoral Program of Higher Education with grant No. 20104104110001,
and  by the Project of Knowledge
Innovation Program (PKIP) of Chinese Academy of Sciences under grant
No. KJCX2.YW.W10.

\end{document}